\documentstyle[12pt,aasms4]{article}
\font\smfont=cmr7
\font\smtfont=cmr5
\font\smttfont=cmr5

\newfam\smfam

\textfont\smfam=\smfont
\scriptfont\smfam=\smtfont
\scriptscriptfont\smfam=\smttfont

\def\spose#1{\hbox to 0pt{#1\hss}}

\def\kms{\ifmmode {\rm\,km\,s^{-1}}\else
    ${\rm\,km\,s^{-1}}$\fi}
\def\kmsMpc{\ifmmode {\rm\,km\,s^{-1}\,Mpc^{-1}}\else
    ${\rm\,km\,s^{-1}\,Mpc^{-1}}$\fi}

\def\msun{\ifmmode {\rm\,M_\odot}\else ${\rm\,M_\odot}$\fi}
\def\Msun{\ifmmode {\rm\,M_\odot}\else ${\rm\,M_\odot}$\fi}
\def\lsun{\ifmmode {\rm\,L_\odot}\else ${\rm\,L_\odot}$\fi}
\def\Lsun{\ifmmode {\rm\,L_\odot}\else ${\rm\,L_\odot}$\fi}
\def\rsun{\ifmmode {\rm\,R_\odot}\else ${\rm\,R_\odot}$\fi}
\def\Rsun{\ifmmode {\rm\,R_\odot}\else ${\rm\,R_\odot}$\fi}

\def\cm{{\rm\,cm}}
\def\cm3{\ifmmode {\rm\,cm^{-3}}\else ${\rm\,cm^{-3}}$\fi}

\def\ergps{\ifmmode {\rm\,erg\,s^{-1}}\else ${\rm\,erg\,s^{-1}}$\fi}
\def\ergpscm2{\ifmmode {\rm\,erg\,s^{-1}\,cm^{-2}}\else
    ${\rm\,erg\,s^{-1}\,cm^{-2}}$\fi}

\def\deg{\ifmmode {^{\circ}}\else {$^\circ$}\fi}
\def\degr{\ifmmode {^{\circ}}\else {$^\circ$}\fi}
\def\degs{\ifmmode {^{\circ}}\else {$^\circ$}\fi}

\def\etal{{\it et al.~}}

\def\h3Mpc{h^{-3}{\rm Mpc}^3}
\def\Ho{\ifmmode {\rm\,H_\circ}\else ${\rm\,H_\circ}$\fi}
\def\hnot{\ifmmode {\rm\,H_\circ}\else ${\rm\,H_\circ}$\fi}
\def\h0{\ifmmode {\rm\,H_\circ}\else ${\rm\,H_\circ}$\fi}
\def\hnotunit{\ifmmode {\rm\,km\,s^{-1}\,Mpc^{-1}}\else
    ${\rm\,km\,s^{-1}\,Mpc^{-1}}$\fi}

\def\qnot{\ifmmode {\rm\,q_\circ}\else ${\rm q_\circ}$\fi}
\def\q0{\ifmmode {\rm\,q_\circ}\else ${\rm q_\circ}$\fi}

\def\arcsec{\ifmmode {^{\prime\prime}}\else $^{\prime\prime}$\fi}
\def\asec{\ifmmode {^{\prime\prime}}\else $^{\prime\prime}$\fi}
\def\arcmin{\ifmmode {^{\prime}}\else $^{\prime}$\fi}
\def\amin{\ifmmode {^{\prime}}\else $^{\prime}$\fi}

\def\secper{\ifmmode \rlap.{^{s}}\else $\rlap{.}{^{s}} $\fi}
\def\minper{\ifmmode \rlap.{^{m}}\else $\rlap{.}{^m} $\fi}
\def\magper{\ifmmode \rlap.{^{m}}\else $\rlap{.}{^m} $\fi}
\def\arcsper{\ifmmode \rlap.{^{\prime\prime}}\else
    $\rlap.{^{\prime\prime}}$\fi}
\def\arcmper{\ifmmode \rlap.{^{\prime}}\else
    $\rlap.{^{\prime}}$\fi}
\def\spose#1{\hbox to 0pt{#1\hss}}
\def\simlt{\mathrel{\spose{\lower 3pt\hbox{$\mathchar"218$}}
 \raise 2.0pt\hbox{$\mathchar"13C$}}}
\def\simgt{\mathrel{\spose{\lower 3pt\hbox{$\mathchar"218$}}
     \raise 2.0pt\hbox{$\mathchar"13E$}}}

\def\refindent{\par\noindent\parskip=2pt\hangindent=3pc\hangafter=1 }

\begin{document}
\title{Radio Observations of 4079 Quasars}
\author{Otto B. Bischof }
\affil{Physics Department, University of California, Davis, CA 95616}
\affil{and}
\affil{Institute of Geophysics \& Planetary Physics, LLNL, Livermore, CA 94550}
\affil{obischof@igpp.llnl.gov}
\author{Robert H. Becker}
\affil{Physics Department, University of California, Davis, CA 95616}
\affil{and}
\affil{Institute of Geophysics \& Planetary Physics, LLNL, Livermore, CA 94550}
\affil{bob@igpp.llnl.gov}
\medskip
\centerline {Version: 10 Feb 1997 }

\begin{abstract}
Using the NVSS radio catalog, we have searched for
radio emission from 4079 quasars taken from the 1996
version of the Veron-Cetty and Veron (1996) quasar catalog. 
The comparison resulted in
the positive detection of radio emission from 799 quasars.
Of these, 168  are new radio detections.       
Examination of the radio luminosities shows a dramatic
increase in the fraction of radio-loud quasars from the 
current epoch to z=0.5 and a gradual decline  beyond
z=1.0. Inspection of the radio-loud fraction as a function of
M$_{\rm B}$ shows little dependence fainter than  M$_{\rm B}$=$-$29.5.

\end{abstract}

\section{Introduction}
Although quasars were originally discovered as a result of their
radio emission, it is now clear that a majority of quasars
are 'radio-quiet'.  It is common to classify quasars as
radio-loud or radio-quiet either based on their radio luminosity
or on the ratio of their radio to optical luminosities.
The latter ratio is known to span a range of 5 orders of magnitude.
To date there is no generally accepted explanation
for the difference between radio-loud and radio-quiet quasars.

One avenue of research has been to search for correlations between
the fraction of radio-loud quasars (F$_{\rm RLQ}$) with other observables.
It is still an open question whether radio-loud and radio-quiet quasars
are two distinct classes of objects or represent a continuous distribution of
radio to optical power. Previous studies have attempted
to define unbiased  samples
of quasars to address the nature of radio-loud quasars. These samples have been
drawn from optically and x-ray selected quasar surveys. Their main drawback
has been the relatively small size of the samples. In contrast, this study
emphasizes quantity at the risk of introducing unrecognized biases.
That is to say, in this work we will examine the radio emission from
all known quasars over a large region of sky independent of 
how the individual quasars were discovered. This has one key
advantage, a very large sample, which optimistically may compensate
for the diversity of the sample. 

In a recent study, Hooper \etal (1996) presented a sample of 61 VLA radio
detections among 359 optically selected quasars from the 
Large Bright Quasar Survey (LBQS) (Hewett \etal 1995). They computed 
F$_{\rm RLQ}$ as a function of redshift and absolute magnitude and found that 
it was only weakly dependent on either quantity. In this paper we present a 
similar study for a much larger sample of quasars. Using the
National Radio Astronomy Observatory Very Large Array Sky Survey (NVSS) 
(Condon 1995), we have searched for radio emission from 4079 quasars
(47 \% of the quasars currently known) in the latest edition of the
Veron catalog (Veron-Cetty and Veron 1996, hereafter VV) which lie
within the area of sky covered by the NVSS.

\section{Radio/Optical Comparisons}

This exercise is possible because of new
radio surveys which are becoming available.  In particular,
the VLA is currently being used to survey 75 \% of the sky at 1400 MHz.
The survey, formally known as the NVSS, has an angular resolution
of $\sim$ 45 arcsec and a 5 sigma sensitivity of $\sim$ 2.4 mJy.
To date, data for
approximately one third of the eventual sky coverage has been released.

Our methodology is straightfoward. We compare the positions of
quasars in VV to the positions of radio sources in the NVSS
catalog and calculate the separation between each quasar and the
nearest radio source. We estimate the chance coincidence rate
by repeating the experiment with a false quasar catalog generated by
shifting the positions in VV by 1 degree. (Note that the chance coincidence
rate is in agreement with expectations for a random distribution.)
The spatial distribution of quasars that fall within the NVSS survey as of
May 1, 1996 is shown in Figure 1. The obvious nonuniform distribution of 
quasars on the sky is a direct result of inhomogeneous sky coverage of surveys.
In Figure 2, we show a histogram of the
angular separations between the known quasars and the nearest radio source
both for VV as well as the false quasar catalog.
We conclude that matches within 10 arcsec are 98 \% real while
matches beyond 35 arcsec are almost totally chance coincidences.
(Note that 10 arcsec is approximately equal to the radius of the
90 \% confidence error circle for a 2.4mJy source). The 149 
matches between 10-35 arcsec are a mixture of real (60 \%) and chance (40 \%)
coincidences and will not be included in this study even though they constitute
10 \% of the real matches. In summary,
we find 799 radio detections of quasars, including 168 new detections which
are listed in Table 1, 3280 nondetections
for which we assign an upper limit of 2.4 mJy, and 149 ambiguous matches
which we discard. There has been no attempt to identify which of the quasars
were originally discovered as a result of their radio emission.

\section{Results}

In Figure 3, we show a histogram of the 1400 MHz flux densities up to
2 Jy of the 799 quasars with counterparts in the NVSS.
The shaded histogram depicts the same for the 168 quasars detected
for the first time here. It is apparent that there are very few new
detections above 100 mJy. 

Using the radio flux densities (or upper limits) in conjunction with the
redshifts as given by VV, we have calculated a radio luminosity (or upper
limit) for each quasar using the equations from Weedman (1986), where we have
assumed a radio spectral index of $-$0.5, a Hubble constant 
of H$_0$= 50\hnotunit\ and q$_0$= 0.5. Furthermore, using the same 
spectral index we have adjusted the
radio luminosity to 8.4 GHz to facilitate comparisons to previously
published results(e.g. Hooper \etal 1996).  In Figures 4 and 5, we
show the distribution of L$_{\rm R}$ as a function of redshift z and M$_{\rm B}$
(as given in VV with a correction to q$_0$=0.5). 
Using 10$^{32}$ ergs/sec/Hz as the division
between radio-loud and radio-quiet, it is clear that the NVSS detects most
of the radio-loud quasars out to a redshift of 1.5 while missing
most of the radio-quiet quasars. Of the 3280 upper limits, 1799 fall above
10$^{32}$ ergs/sec/Hz.

It may be more informative to consider the ratio of radio to optical
luminosity L$_{\rm R}$/L$_{\rm O}$ in defining radio loud quasars. 
In Figures 6 and 7,
we plot log(L$_{\rm R}$/L$_{\rm O}$) against z and M$_{\rm B}$.  
In this context, the divide
between radio-loud and radio-quiet is often taken to be 1.0. There are 
only 139 upper limits which lie above this divide. Lastly, using 1.0 as 
the divide between radio-loud and radio-quiet, we calculate F$_{\rm RLQ}$
as a function of M$_{\rm B}$ and z (see Figures 8 and 9) where we have
not included upper limits above 1.0 in the calculation.
The dependence of F$_{\rm RLQ}$ on M$_{\rm B}$ is minimal,
remaining nearly constant at $\sim$ 0.16 between M$_{\rm B}$
of $-$23 to $-$29. At M$_{\rm B}$=$-$30, F$_{\rm RLQ}$ may decline to 
below 0.10. 
The case is quite
different for the dependence on redshift. F$_{\rm RLQ}$ rises dramatically
between z=0 and z=0.5 from an initial value of 0.05 up to 0.25. Beyond z=1.0, 
F$_{\rm RLQ}$ appears to slowly decrease to 0.12 at z=2 and then remain 
approximately constant out to z=4. These results are similar to the z 
dependence reported by Hooper \etal (1995,1996) on a much smaller sample 
from the LBQS.

\section{Discussion}

Considerable effort has been expended over the past several decades
to understand the underlying difference between radio-loud and
radio-quiet quasars. These studies have recently focussed on 
two specific questions; namely, do the radio properties of
quasars evolve with time and do they correlate with optical
luminosity? Unfortunately the answers to these questions seem to
differ from quasar sample to quasar sample. For example,
Hooper \etal (1996) found that F$_{\rm RLQ}$ was
nearly constant at 0.10 over a large range of absolute B magnitude.
In contrast they pointed out that a similar analysis of the Palomar Green
sample (Schmidt and Green 1983; Kellermann \etal 1989)
revealed strong variations in
F$_{\rm RLQ}$ with absolute B magnitude. Hooper \etal (1996) were at a loss
to explain the difference.

Our comparison between the VV catalog and the NVSS has yielded radio
data on a quasar sample which is an order of magnitude larger than
previous samples. In general, the radio properties of this sample are
very similar to those of the LBQS. In particular, F$_{\rm RLQ}$ for 
this new large sample is also largely independent of absolute B magnitude
as shown in Figure 8. Likewise, in agreement with the LBQS sample, we see
a peak in F$_{\rm RLQ}$ between z of 0.5 to 1.0. However,
with this larger sample, we can study the dependence on z with much
higher resolution and as a result, find a significantly stronger dependence
on z. In particular, we find that F$_{\rm RLQ}$ 
is a factor of five lower today than it was at z=0.5. While there has been
considerable discussion in the literature on the differential
evolution of radio-loud quasars at very high redshifts, our results
suggest that differential evolution of radio-loud quasars has been most
important over the last several billion years.

As we noted at the outset of this paper, to obtain a larger sample we
sacrificed the uniformity of the sample. It is possible that the extreme
differential evolution we observe between z of 0 and 0.5 is a consequence
of various selection effects. For example, the nearby quasar population
has a higher proportion of low luminosity objects. This in principle
could bias the dependence of radio-loud fraction on redshift. However
we do not see a dependence of radio-loud fraction on optical luminosity.
While we cannot rule out some unforseen bias, we are encouraged that the LBQS
sample also shows a peak at z of 1, albeit with less statistical significance
and resolution.

This type of study and the conclusions drawn from it certainly should be
used carefully in light of the heterogenous nature of the quasar sample.
This is particularly true in regions of parameter space where the sample is
very sparse, such as at very high redshift or luminosity. However, selection
effects are probably less important at lower redshifts where quasars are
well-sampled and easier to detect. In particular, it is hard to see how
radio-loud quasars could be selected against at low redshift, so we think
it very likely that this absence at low redshift is a real effect.

\centerline{\bf Acknowledgements}
\par
The authors thank Chris Impey for many useful comments.
This work was supported in part under the auspices of the Department of Energy
by Lawrence Livermore National Laboratory under contract W-7405-ENG-48 and by
the Institute of Geophysics and Planetary Physics.

\newpage
\centerline{\bf Figure Captions}
\figcaption{Location of VV quasars falling within the area covered by the NVSS.
Right ascension increases from left to right in the figure starting at 0 hours. 
Declination varies in degrees from $-$90 to +90 as shown.}
\figcaption{A histogram of the separations between VV quasars and NVSS radio
sources as well as an estimate of the chance coincidence rate using a false
quasar catalog generated by shifting VV positions by 1 degree.}
\figcaption{A histogram of radio flux densities for the 799 detections of
VV quasars. The shaded histogram is the subset of new detections.}
\figcaption{The distribution of radio luminosity corrected to 8.4 GHz vs. redshift. The triangles are positive detections and the dots are upper limits.}
\figcaption{The distribution of radio luminosity at 8.4 GHz versus
absolute B magnitude. The triangles are positive detections and the dots 
are upper limits.}
\figcaption{Distribution of ratio of radio to optical luminosity versus 
redshift. Upper limits are dots and positive detections are triangles.}
\figcaption{Distribution of ratio of radio to optical luminosity versus 
absolute B magnitude. Upper limits are dots and positive detections are 
triangles.}
\figcaption{Radio loud fraction as function of absolute B magnitude. Error bars
are plus or minus one standard deviation. The error is for the standard
deviation of a random binomial distribution in the radio loud fraction.}
\figcaption{Radio loud fraction as function of redshift. The error is computed
the same way as in the previous plot.}
\newpage
\begin{planotable}{lccccc}
\tablewidth{0pc}
\tablecaption{New Quasar Radio Detections}
\tablenum{1}
\tablehead{
\colhead{R.A.(2000)}&
\colhead{Dec.(2000)} & \colhead{$z$} & \colhead{S$_{\rm 1400}$(mJY)} & \colhead{M$_{\rm B}$} & \colhead{logL$_{\rm 8400}$} }
\startdata
00\phantom{q}00\phantom{q}41.9 & -02\phantom{q}29\phantom{q}23 & 1.07 &    7.2 &
 -26.3 & 32.1 \nl
00\phantom{q}01\phantom{q}50.0 & -01\phantom{q}59\phantom{q}40 & 2.82 &   37.1 &
 -27.4 & 33.7 \nl
00\phantom{q}05\phantom{q}07.0 & -01\phantom{q}32\phantom{q}46 & 1.71 &   73.1 &
 -25.9 & 33.5 \nl
00\phantom{q}07\phantom{q}27.1 &  02\phantom{q}41\phantom{q}12 & 0.30 &    8.0 &
 -23.5 & 31.1 \nl
00\phantom{q}27\phantom{q}17.3 &  00\phantom{q}37\phantom{q}23 & 1.23 &    4.6 &
 -26.1 & 32.1 \nl
00\phantom{q}31\phantom{q}36.5 & -01\phantom{q}36\phantom{q}22 & 2.38 &   16.6 &
 -26.7 & 33.2 \nl
00\phantom{q}39\phantom{q}42.4 & -39\phantom{q}29\phantom{q}14 & 1.46 &    4.6 &
 -25.9 & 32.2 \nl
00\phantom{q}39\phantom{q}42.5 & -35\phantom{q}28\phantom{q}02 & 0.84 &   92.2 &
 -25.2 & 33.0 \nl
00\phantom{q}41\phantom{q}50.6 & -38\phantom{q}20\phantom{q}09 & 1.74 &   71.2 &
 -24.8 & 33.5 \nl
00\phantom{q}41\phantom{q}59.9 & -39\phantom{q}21\phantom{q}47 & 1.85 &    4.9 &
 -26.7 & 32.4 \nl
00\phantom{q}42\phantom{q}16.5 & -02\phantom{q}54\phantom{q}22 & 2.74 &   31.6 &
 -27.3 & 33.6 \nl
00\phantom{q}42\phantom{q}40.7 & -33\phantom{q}39\phantom{q}20 & 2.06 &    6.4 &
 -25.8 & 32.6 \nl
00\phantom{q}42\phantom{q}59.9 & -30\phantom{q}07\phantom{q}42 & 0.61 &   39.1 &
 -24.7 & 32.4 \nl
00\phantom{q}43\phantom{q}22.4 & -26\phantom{q}39\phantom{q}05 & 1.00 &   80.1 &
 -26.4 & 33.1 \nl
00\phantom{q}43\phantom{q}51.3 & -28\phantom{q}27\phantom{q}41 & 0.84 &   55.2 &
 -25.1 & 32.8 \nl
00\phantom{q}48\phantom{q}10.3 & -40\phantom{q}05\phantom{q}34 & 1.64 &    8.1 &
 -26.2 & 32.6 \nl
00\phantom{q}49\phantom{q}40.8 & -26\phantom{q}32\phantom{q}59 & 3.16 &   39.0 &
 -25.9 & 33.8 \nl
00\phantom{q}50\phantom{q}40.9 & -25\phantom{q}41\phantom{q}24 & 0.78 &   43.7 &
 -24.5 & 32.7 \nl
00\phantom{q}51\phantom{q}30.4 &  00\phantom{q}41\phantom{q}50 & 1.19 &   16.3 &
 -25.8 & 32.6 \nl
00\phantom{q}51\phantom{q}53.6 & -27\phantom{q}56\phantom{q}18 & 1.15 &    6.8 &
 -23.8 & 32.2 \nl
00\phantom{q}52\phantom{q}05.5 &  00\phantom{q}35\phantom{q}38 & 0.40 &  135.3 &
 -25.3 & 32.6 \nl
00\phantom{q}52\phantom{q}45.4 & -26\phantom{q}24\phantom{q}23 & 2.22 &    5.6 &
 -25.0 & 32.6 \nl
00\phantom{q}52\phantom{q}52.4 & -28\phantom{q}25\phantom{q}55 & 1.65 &  226.6 &
 -26.1 & 34.0 \nl
00\phantom{q}54\phantom{q}04.0 & -27\phantom{q}10\phantom{q}10 & 0.69 &    7.0 &
 -23.9 & 31.8 \nl
00\phantom{q}54\phantom{q}24.9 & -25\phantom{q}24\phantom{q}00 & 1.95 &  112.5 &
 -24.7 & 33.8 \nl
00\phantom{q}54\phantom{q}41.2 &  00\phantom{q}01\phantom{q}10 & 0.65 &    4.4 &
 -25.0 & 31.5 \nl
00\phantom{q}54\phantom{q}45.4 & -38\phantom{q}44\phantom{q}15 & 3.18 &   29.0 &
 -26.9 & 33.7 \nl
00\phantom{q}55\phantom{q}19.8 & -29\phantom{q}59\phantom{q}33 & 2.10 &   40.1 &
 -25.3 & 33.5 \nl
00\phantom{q}55\phantom{q}35.1 &  01\phantom{q}40\phantom{q}35 & 0.44 &   26.8 &
 -24.2 & 32.0 \nl
00\phantom{q}55\phantom{q}42.8 & -27\phantom{q}23\phantom{q}19 & 1.54 &    9.9 &
 -25.1 & 32.6 \nl
00\phantom{q}56\phantom{q}04.3 & -26\phantom{q}52\phantom{q}57 & 1.04 &   21.7 &
 -25.2 & 32.6 \nl
00\phantom{q}58\phantom{q}01.8 &  15\phantom{q}53\phantom{q}15 & 1.26 &    6.1 &
 -26.3 & 32.2 \nl
00\phantom{q}58\phantom{q}41.2 & -39\phantom{q}08\phantom{q}42 & 1.41 &    8.6 &
 -26.8 & 32.5 \nl
00\phantom{q}59\phantom{q}17.6 &  01\phantom{q}42\phantom{q}06 & 3.15 &    6.5 &
 -27.8 & 33.0 \nl
00\phantom{q}59\phantom{q}58.1 & -39\phantom{q}35\phantom{q}56 & 2.04 &   10.2 &
 -25.7 & 32.8 \nl
01\phantom{q}00\phantom{q}12.3 & -27\phantom{q}08\phantom{q}53 & 3.52 &  210.9 &
 -27.4 & 34.6 \nl
01\phantom{q}00\phantom{q}59.8 & -28\phantom{q}51\phantom{q}15 & 0.87 &    8.8 &
 -25.4 & 32.1 \nl
01\phantom{q}02\phantom{q}00.6 & -30\phantom{q}18\phantom{q}27 & 1.03 &   78.7 &
 -26.6 & 33.2 \nl
01\phantom{q}02\phantom{q}59.8 & -27\phantom{q}23\phantom{q}43 & 1.42 &    3.8 &
 -25.2 & 32.1 \nl
01\phantom{q}03\phantom{q}29.4 &  00\phantom{q}40\phantom{q}54 & 1.44 &  117.9 &
 -25.5 & 33.6 \nl
01\phantom{q}04\phantom{q}03.5 & -29\phantom{q}43\phantom{q}53 & 1.32 &   13.1 &
 -24.1 & 32.6 \nl
01\phantom{q}09\phantom{q}28.8 & -34\phantom{q}05\phantom{q}18 & 0.83 &   26.8 &
 -23.3 & 32.5 \nl
01\phantom{q}09\phantom{q}52.0 & -35\phantom{q}25\phantom{q}56 & 2.19 &    2.8 &
 -25.0 & 32.3 \nl
01\phantom{q}11\phantom{q}08.1 & -35\phantom{q}35\phantom{q}17 & 2.36 &   15.0 &
 -26.4 & 33.1 \nl
01\phantom{q}11\phantom{q}19.5 & -33\phantom{q}08\phantom{q}22 & 2.31 &    3.1 &
 -26.0 & 32.4 \nl
01\phantom{q}15\phantom{q}47.7 & -33\phantom{q}07\phantom{q}15 & 1.91 &   80.1 &
 -24.3 & 33.7 \nl
01\phantom{q}16\phantom{q}36.9 & -34\phantom{q}57\phantom{q}01 & 1.96 &  107.5 &
 -25.7 & 33.8 \nl
01\phantom{q}22\phantom{q}04.4 & -33\phantom{q}10\phantom{q}12 & 1.11 &   17.9 &
 -26.7 & 32.6 \nl
01\phantom{q}26\phantom{q}55.8 & -32\phantom{q}08\phantom{q}10 & 2.20 &    4.5 &
 -27.5 & 32.5 \nl
01\phantom{q}27\phantom{q}44.0 & -34\phantom{q}57\phantom{q}54 & 2.16 &   18.8 &
 -26.5 & 33.2 \nl
01\phantom{q}34\phantom{q}05.1 & -39\phantom{q}32\phantom{q}15 & 1.89 &   12.1 &
 -24.8 & 32.8 \nl
01\phantom{q}34\phantom{q}38.6 & -19\phantom{q}32\phantom{q}07 & 3.13 &   47.4 &
 -28.4 & 33.9 \nl
01\phantom{q}35\phantom{q}05.7 & -07\phantom{q}20\phantom{q}29 & 1.64 &    6.8 &
 -26.1 & 32.5 \nl
01\phantom{q}35\phantom{q}35.7 & -38\phantom{q}54\phantom{q}25 & 2.08 &  182.0 &
 -25.3 & 34.1 \nl
01\phantom{q}35\phantom{q}55.5 &  00\phantom{q}39\phantom{q}24 & 1.46 &    4.7 &
 -24.2 & 32.2 \nl
01\phantom{q}42\phantom{q}54.7 & -30\phantom{q}23\phantom{q}45 & 3.12 &    5.0 &
 -28.4 & 32.9 \nl
01\phantom{q}42\phantom{q}54.8 & -38\phantom{q}56\phantom{q}46 & 2.21 &    4.0 &
 -25.6 & 32.5 \nl
01\phantom{q}52\phantom{q}27.3 & -20\phantom{q}01\phantom{q}07 & 2.15 &   26.9 &
 -27.7 & 33.3 \nl
01\phantom{q}52\phantom{q}59.1 & -01\phantom{q}29\phantom{q}40 & 2.02 &   47.7 &
 -26.8 & 33.5 \nl
01\phantom{q}57\phantom{q}08.6 &  01\phantom{q}38\phantom{q}24 & 2.13 &    4.8 &
 -26.7 & 32.5 \nl
01\phantom{q}59\phantom{q}21.0 &  19\phantom{q}00\phantom{q}32 & 2.61 &   21.9 &
 -26.1 & 33.4 \nl
02\phantom{q}16\phantom{q}12.3 & -01\phantom{q}05\phantom{q}19 & 2.15 &  144.0 &
 -27.0 & 34.0 \nl
02\phantom{q}42\phantom{q}05.2 & -23\phantom{q}44\phantom{q}53 & 2.63 &   29.5 &
 -26.4 & 33.5 \nl
02\phantom{q}42\phantom{q}40.3 &  00\phantom{q}57\phantom{q}27 & 0.57 &    8.3 &
 -25.8 & 31.7 \nl
02\phantom{q}44\phantom{q}50.5 & -22\phantom{q}16\phantom{q}01 & 2.20 &   24.0 &
 -24.9 & 33.3 \nl
02\phantom{q}45\phantom{q}47.8 & -31\phantom{q}37\phantom{q}25 & 1.88 &   13.6 &
 -26.3 & 32.9 \nl
02\phantom{q}47\phantom{q}39.1 & -00\phantom{q}52\phantom{q}22 & 2.12 &    7.9 &
 -25.0 & 32.8 \nl
02\phantom{q}48\phantom{q}06.6 & -29\phantom{q}36\phantom{q}36 & 1.86 &    3.5 &
 -25.3 & 32.3 \nl
02\phantom{q}50\phantom{q}48.6 &  00\phantom{q}02\phantom{q}08 & 0.77 &  112.7 &
 -24.1 & 33.1 \nl
02\phantom{q}51\phantom{q}46.8 &  15\phantom{q}50\phantom{q}13 & 0.49 &   30.4 &
 -23.1 & 32.1 \nl
02\phantom{q}51\phantom{q}56.2 &  00\phantom{q}57\phantom{q}06 & 0.47 &    7.5 &
 -23.4 & 31.5 \nl
02\phantom{q}56\phantom{q}58.5 &  00\phantom{q}54\phantom{q}47 & 1.11 &   13.2 &
 -24.7 & 32.4 \nl
03\phantom{q}00\phantom{q}29.8 &  02\phantom{q}40\phantom{q}50 & 0.12 &    6.8 &
 -23.0 & 30.2 \nl
03\phantom{q}01\phantom{q}38.2 &  01\phantom{q}38\phantom{q}17 & 1.58 &    8.1 &
 -25.4 & 32.5 \nl
03\phantom{q}05\phantom{q}48.9 & -02\phantom{q}40\phantom{q}13 & 1.27 &   18.6 &
 -25.6 & 32.7 \nl
03\phantom{q}10\phantom{q}02.0 & -37\phantom{q}23\phantom{q}38 & 0.40 &   15.5 &
 -24.1 & 31.6 \nl
03\phantom{q}11\phantom{q}39.8 & -40\phantom{q}08\phantom{q}42 & 1.73 &    9.5 &
 -26.1 & 32.7 \nl
03\phantom{q}16\phantom{q}31.3 &  01\phantom{q}37\phantom{q}30 & 0.96 &   11.5 &
 -24.7 & 32.3 \nl
04\phantom{q}07\phantom{q}28.8 &  17\phantom{q}50\phantom{q}52 & 1.71 &  551.6 &
 -25.4 & 34.4 \nl
04\phantom{q}22\phantom{q}41.8 &  00\phantom{q}30\phantom{q}20 & 2.92 &   20.6 &
 -27.2 & 33.4 \nl
04\phantom{q}29\phantom{q}35.5 & -22\phantom{q}41\phantom{q}03 & 0.27 &   15.8 &
 -24.1 & 31.3 \nl
04\phantom{q}38\phantom{q}48.1 & -36\phantom{q}35\phantom{q}09 & 0.38 &   10.6 &
 -24.2 & 31.4 \nl
04\phantom{q}41\phantom{q}22.6 & -27\phantom{q}08\phantom{q}20 & 0.08 &   34.8 &
 -23.0 & 30.6 \nl
04\phantom{q}43\phantom{q}14.3 & -36\phantom{q}46\phantom{q}24 & 0.68 &    4.7 &
 -25.6 & 31.6 \nl
04\phantom{q}44\phantom{q}15.2 & -30\phantom{q}04\phantom{q}52 & 2.40 &   10.1 &
 -27.4 & 33.0 \nl
04\phantom{q}49\phantom{q}21.1 &  07\phantom{q}29\phantom{q}10 & 1.46 &   61.8 &
 -27.5 & 33.3 \nl
04\phantom{q}52\phantom{q}30.1 & -29\phantom{q}53\phantom{q}35 & 0.29 &   10.0 &
 -25.1 & 31.2 \nl
05\phantom{q}04\phantom{q}55.3 & -27\phantom{q}23\phantom{q}11 & 1.14 &  183.4 &
 -26.4 & 33.6 \nl
05\phantom{q}58\phantom{q}13.5 & -36\phantom{q}19\phantom{q}48 & 2.22 &    3.1 &
 -27.0 & 32.4 \nl
06\phantom{q}06\phantom{q}07.2 & -34\phantom{q}47\phantom{q}40 & 2.28 &  280.4 &
 -26.5 & 34.4 \nl
07\phantom{q}07\phantom{q}13.2 &  64\phantom{q}35\phantom{q}59 & 0.08 &    3.2 &
 -23.1 & 29.5 \nl
07\phantom{q}51\phantom{q}00.7 &  03\phantom{q}20\phantom{q}41 & 0.10 &   12.0 &
 -23.6 & 30.3 \nl
07\phantom{q}58\phantom{q}19.8 &  42\phantom{q}19\phantom{q}35 & 0.21 &    4.0 &
 -24.1 & 30.5 \nl
08\phantom{q}15\phantom{q}20.6 &  27\phantom{q}36\phantom{q}18 & 0.91 &    6.0 &
 -27.4 & 31.9 \nl
08\phantom{q}36\phantom{q}58.9 &  44\phantom{q}26\phantom{q}02 & 0.25 &    7.9 &
 -25.3 & 30.9 \nl
08\phantom{q}38\phantom{q}40.5 &  47\phantom{q}34\phantom{q}10 & 0.70 &   29.6 &
 -23.7 & 32.4 \nl
08\phantom{q}41\phantom{q}18.0 &  35\phantom{q}44\phantom{q}38 & 1.77 &    6.7 &
 -28.7 & 32.5 \nl
08\phantom{q}42\phantom{q}15.2 &  45\phantom{q}25\phantom{q}44 & 1.41 &   75.4 &
 -26.9 & 33.4 \nl
08\phantom{q}47\phantom{q}56.4 &  31\phantom{q}47\phantom{q}58 & 1.83 &   29.0 &
 -26.8 & 33.2 \nl
08\phantom{q}48\phantom{q}47.8 &  14\phantom{q}20\phantom{q}56 & 1.69 &   44.3 &
 -26.5 & 33.3 \nl
08\phantom{q}55\phantom{q}16.1 &  56\phantom{q}16\phantom{q}56 & 0.44 &    5.6 &
 -23.3 & 31.3 \nl
08\phantom{q}56\phantom{q}58.8 &  51\phantom{q}20\phantom{q}45 & 2.31 &    8.6 &
 -25.8 & 32.9 \nl
08\phantom{q}57\phantom{q}06.3 &  19\phantom{q}08\phantom{q}54 & 0.33 &    5.2 &
 -23.8 & 31.0 \nl
09\phantom{q}06\phantom{q}03.6 &  19\phantom{q}41\phantom{q}44 & 1.21 &   17.4 &
 -27.2 & 32.6 \nl
09\phantom{q}28\phantom{q}37.9 &  60\phantom{q}25\phantom{q}21 & 0.30 &   67.8 &
 -23.1 & 32.0 \nl
09\phantom{q}29\phantom{q}58.4 & -26\phantom{q}00\phantom{q}31 & 2.15 &    2.8 &
 -27.1 & 32.3 \nl
09\phantom{q}32\phantom{q}35.0 &  21\phantom{q}58\phantom{q}30 & 2.31 &    5.9 &
 -25.4 & 32.7 \nl
09\phantom{q}38\phantom{q}57.2 &  42\phantom{q}48\phantom{q}29 & 2.04 &    6.5 &
 -26.2 & 32.6 \nl
09\phantom{q}41\phantom{q}28.8 &  44\phantom{q}47\phantom{q}43 & 0.80 &   69.3 &
 -24.6 & 32.9 \nl
09\phantom{q}47\phantom{q}04.5 &  47\phantom{q}21\phantom{q}43 & 0.54 &    4.0 &
 -24.2 & 31.3 \nl
09\phantom{q}47\phantom{q}56.0 &  53\phantom{q}50\phantom{q}01 & 0.49 &   23.1 &
 -25.2 & 32.0 \nl
09\phantom{q}48\phantom{q}35.9 &  43\phantom{q}23\phantom{q}02 & 1.89 &    2.8 &
 -26.7 & 32.2 \nl
09\phantom{q}52\phantom{q}27.2 &  50\phantom{q}48\phantom{q}51 & 1.55 &   72.3 &
 -26.7 & 33.5 \nl
09\phantom{q}55\phantom{q}38.0 &  33\phantom{q}35\phantom{q}04 & 2.50 &   34.1 &
 -28.5 & 33.5 \nl
10\phantom{q}02\phantom{q}54.6 &  32\phantom{q}40\phantom{q}39 & 0.83 &   11.5 &
 -26.6 & 32.1 \nl
10\phantom{q}17\phantom{q}03.4 &  59\phantom{q}24\phantom{q}29 & 0.85 &  340.5 &
 -25.9 & 33.6 \nl
10\phantom{q}28\phantom{q}09.6 & -26\phantom{q}44\phantom{q}18 & 2.90 &  933.8 &
 -28.3 & 35.1 \nl
10\phantom{q}38\phantom{q}08.1 &  47\phantom{q}31\phantom{q}36 & 2.96 &    8.6 &
 -25.7 & 33.1 \nl
10\phantom{q}53\phantom{q}33.4 & -27\phantom{q}39\phantom{q}07 & 0.16 &   12.9 &
 -23.8 & 30.8 \nl
11\phantom{q}03\phantom{q}25.3 & -26\phantom{q}45\phantom{q}15 & 2.15 &   15.8 &
 -28.9 & 33.1 \nl
11\phantom{q}22\phantom{q}41.5 &  30\phantom{q}35\phantom{q}35 & 1.81 &   10.4 &
 -27.7 & 32.7 \nl
11\phantom{q}27\phantom{q}36.4 &  26\phantom{q}54\phantom{q}50 & 0.38 &    3.2 &
 -24.7 & 30.9 \nl
11\phantom{q}30\phantom{q}04.8 &  41\phantom{q}16\phantom{q}19 & 0.72 &    3.4 &
 -26.0 & 31.5 \nl
11\phantom{q}33\phantom{q}14.8 &  28\phantom{q}12\phantom{q}00 & 0.51 &   12.9 &
 -24.6 & 31.8 \nl
11\phantom{q}41\phantom{q}16.1 &  21\phantom{q}56\phantom{q}22 & 0.06 &    6.1 &
 -23.0 & 29.6 \nl
11\phantom{q}48\phantom{q}00.0 &  26\phantom{q}35\phantom{q}44 & 0.87 &  351.4 &
 -25.3 & 33.7 \nl
11\phantom{q}48\phantom{q}18.8 &  31\phantom{q}54\phantom{q}11 & 0.55 &   94.7 &
 -25.0 & 32.7 \nl
11\phantom{q}52\phantom{q}27.5 &  32\phantom{q}09\phantom{q}59 & 0.37 &   41.7 &
 -23.6 & 32.0 \nl
11\phantom{q}57\phantom{q}34.7 &  73\phantom{q}10\phantom{q}37 & 0.51 &    9.5 &
 -25.7 & 31.6 \nl
12\phantom{q}02\phantom{q}43.3 &  37\phantom{q}35\phantom{q}50 & 0.50 &   16.7 &
 -24.2 & 31.9 \nl
12\phantom{q}12\phantom{q}09.4 &  46\phantom{q}23\phantom{q}46 & 2.29 &    4.8 &
 -25.6 & 32.6 \nl
12\phantom{q}17\phantom{q}21.3 &  30\phantom{q}56\phantom{q}31 & 0.31 &    8.4 &
 -24.2 & 31.1 \nl
12\phantom{q}17\phantom{q}36.6 &  51\phantom{q}55\phantom{q}11 & 2.23 &   85.2 &
 -28.6 & 33.8 \nl
12\phantom{q}19\phantom{q}43.6 &  54\phantom{q}08\phantom{q}35 & 1.66 &   45.0 &
 -26.1 & 33.3 \nl
12\phantom{q}33\phantom{q}22.9 &  47\phantom{q}38\phantom{q}43 & 2.23 &   24.3 &
 -25.0 & 33.3 \nl
12\phantom{q}37\phantom{q}14.4 &  26\phantom{q}19\phantom{q}00 & 2.20 &    3.9 &
 -23.4 & 32.5 \nl
13\phantom{q}10\phantom{q}21.1 &  47\phantom{q}34\phantom{q}03 & 3.58 &    9.2 &
 -24.2 & 33.3 \nl
13\phantom{q}11\phantom{q}04.3 &  29\phantom{q}26\phantom{q}13 & 1.82 &   10.4 &
 -27.4 & 32.8 \nl
13\phantom{q}12\phantom{q}11.0 &  48\phantom{q}09\phantom{q}25 & 0.72 &   45.0 &
 -25.4 & 32.6 \nl
13\phantom{q}18\phantom{q}01.9 &  47\phantom{q}06\phantom{q}27 & 2.59 &    2.8 &
 -27.3 & 32.5 \nl
13\phantom{q}18\phantom{q}50.9 &  26\phantom{q}41\phantom{q}42 & 1.91 &   23.1 &
 -23.8 & 33.1 \nl
13\phantom{q}19\phantom{q}13.8 &  26\phantom{q}47\phantom{q}54 & 2.26 &    8.2 &
 -25.1 & 32.8 \nl
13\phantom{q}22\phantom{q}50.7 &  47\phantom{q}39\phantom{q}37 & 1.11 &    8.1 &
 -25.0 & 32.2 \nl
13\phantom{q}36\phantom{q}02.8 &  27\phantom{q}27\phantom{q}47 & 1.12 &   75.5 &
 -24.5 & 33.2 \nl
13\phantom{q}42\phantom{q}08.3 &  27\phantom{q}09\phantom{q}23 & 1.19 &  270.8 &
 -25.0 & 33.8 \nl
13\phantom{q}44\phantom{q}25.6 &  26\phantom{q}14\phantom{q}10 & 1.18 &   11.2 &
 -24.5 & 32.4 \nl
13\phantom{q}45\phantom{q}43.8 &  26\phantom{q}25\phantom{q}07 & 2.03 &    9.5 &
 -24.6 & 32.8 \nl
13\phantom{q}45\phantom{q}44.5 &  26\phantom{q}25\phantom{q}06 & 2.03 &    9.5 &
 -25.0 & 32.8 \nl
14\phantom{q}23\phantom{q}14.3 &  50\phantom{q}55\phantom{q}39 & 0.28 &  180.5 &
 -24.3 & 32.4 \nl
14\phantom{q}43\phantom{q}00.6 &  52\phantom{q}01\phantom{q}37 & 1.57 & 2550.8 &
 -24.5 & 35.0 \nl
14\phantom{q}49\phantom{q}20.7 &  42\phantom{q}21\phantom{q}02 & 0.18 &  160.1 &
 -23.1 & 31.9 \nl
14\phantom{q}50\phantom{q}05.0 &  46\phantom{q}35\phantom{q}19 & 0.29 &    4.6 &
 -23.1 & 30.8 \nl
15\phantom{q}19\phantom{q}07.5 &  52\phantom{q}06\phantom{q}05 & 0.14 &   11.1 &
 -23.3 & 30.6 \nl
15\phantom{q}20\phantom{q}43.6 &  47\phantom{q}32\phantom{q}49 & 2.81 &   87.5 &
 -26.2 & 34.0 \nl
15\phantom{q}29\phantom{q}07.4 &  56\phantom{q}16\phantom{q}07 & 0.10 &    6.0 &
 -23.0 & 30.0 \nl
15\phantom{q}29\phantom{q}17.6 &  47\phantom{q}38\phantom{q}38 & 3.07 &   55.5 &
 -26.2 & 33.9 \nl
15\phantom{q}30\phantom{q}45.8 &  38\phantom{q}39\phantom{q}53 & 2.02 &    3.0 &
 -28.0 & 32.3 \nl
15\phantom{q}34\phantom{q}57.2 &  58\phantom{q}39\phantom{q}24 & 1.90 &  129.5 &
 -25.9 & 33.9 \nl
15\phantom{q}40\phantom{q}58.6 &  47\phantom{q}38\phantom{q}28 & 2.56 &  230.0 &
 -26.0 & 34.4 \nl
15\phantom{q}56\phantom{q}10.6 &  37\phantom{q}40\phantom{q}40 & 2.66 &    9.2 &
 -26.9 & 33.0 \nl
16\phantom{q}25\phantom{q}48.8 &  26\phantom{q}46\phantom{q}59 & 2.52 &    6.9 &
 -29.6 & 32.8 \nl
16\phantom{q}30\phantom{q}20.8 &  37\phantom{q}56\phantom{q}56 & 0.39 &   23.8 &
 -24.8 & 31.8 \nl
16\phantom{q}32\phantom{q}49.6 &  37\phantom{q}16\phantom{q}31 & 2.94 &    3.2 &
 -27.6 & 32.6 \nl
16\phantom{q}45\phantom{q}01.8 &  39\phantom{q}25\phantom{q}48 & 2.15 &    3.0 &
 -26.0 & 32.3 \nl
16\phantom{q}50\phantom{q}05.5 &  41\phantom{q}40\phantom{q}33 & 0.59 &  234.7 &
 -25.2 & 33.1 \nl
16\phantom{q}56\phantom{q}41.1 &  59\phantom{q}15\phantom{q}39 & 0.50 &   13.0 &
 -23.7 & 31.8 \nl
17\phantom{q}51\phantom{q}05.5 &  26\phantom{q}59\phantom{q}02 & 0.14 &  146.4 &
 -24.5 & 31.7 \nl
18\phantom{q}30\phantom{q}23.3 &  73\phantom{q}13\phantom{q}10 & 0.12 &    9.1 &
 -23.8 & 30.4 \nl

\enddata
\end{planotable}




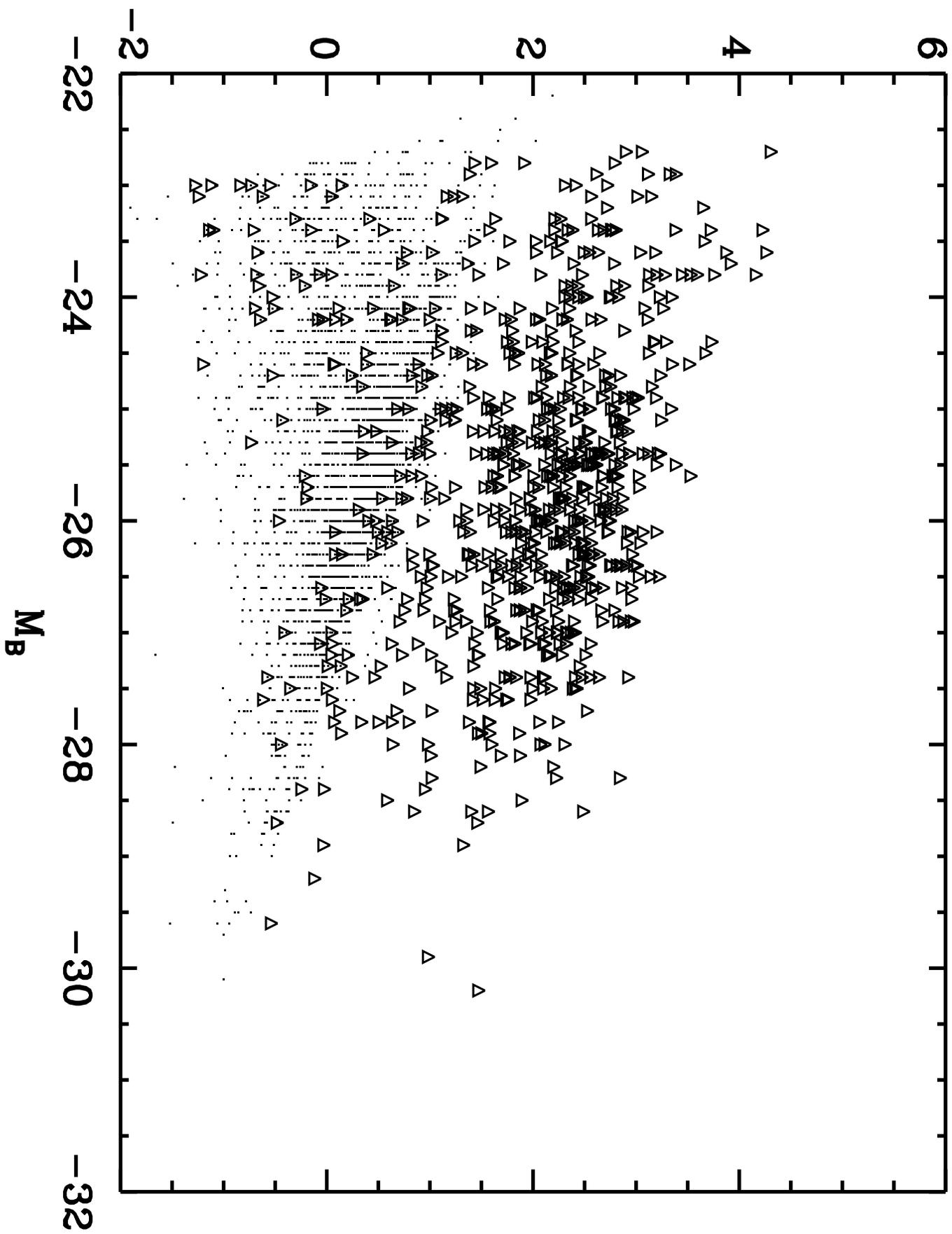
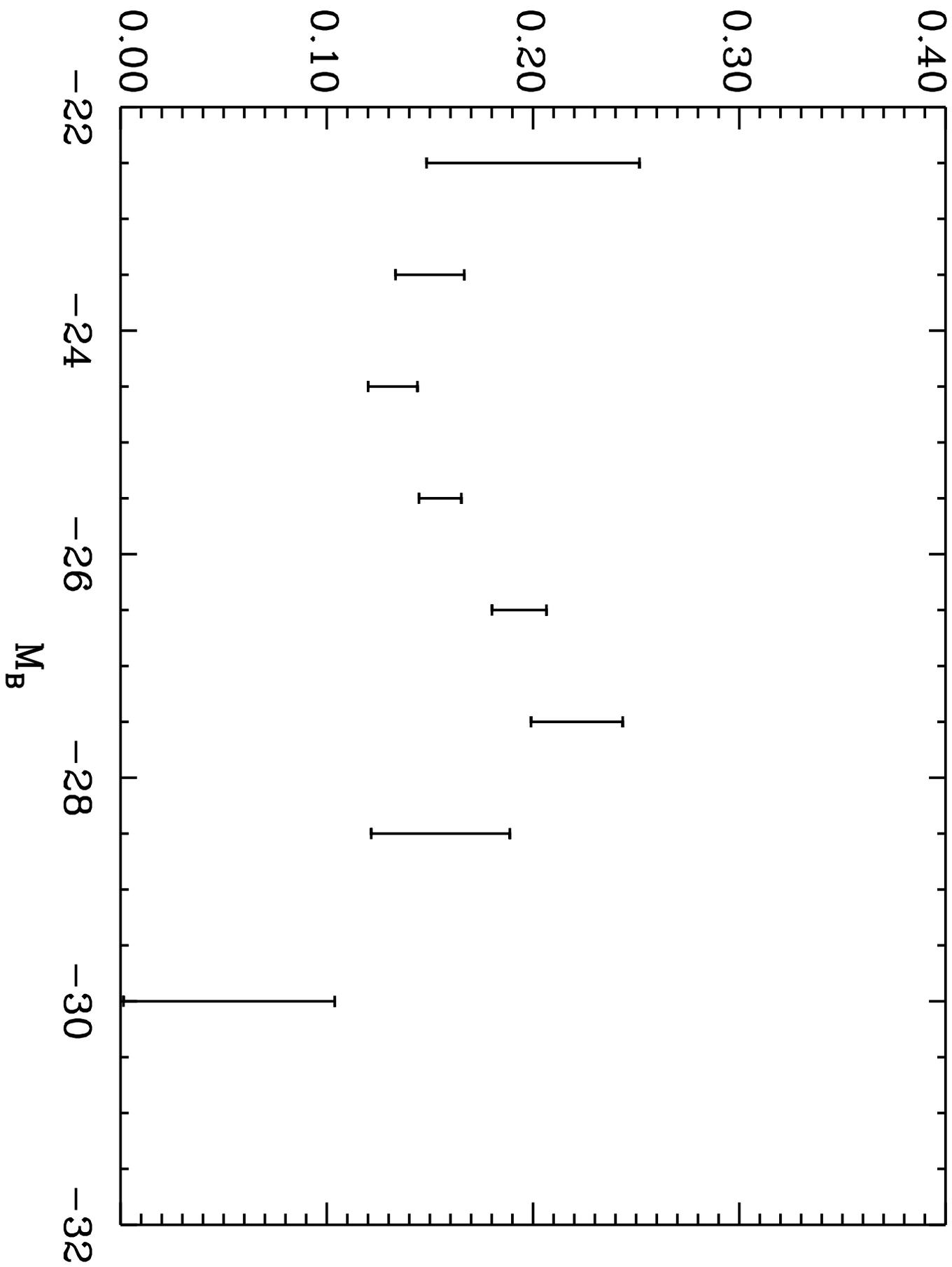
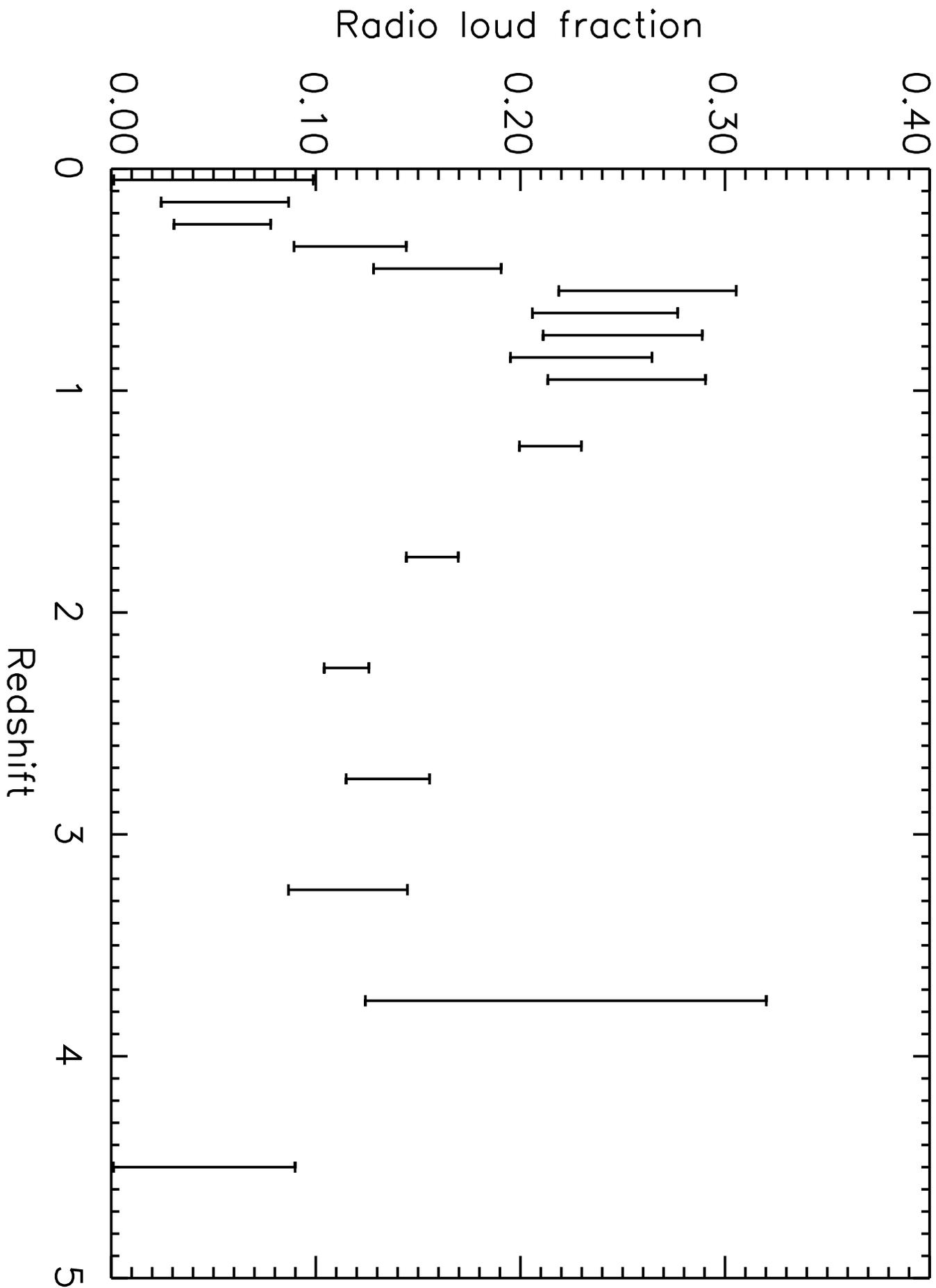
\end{document}